\documentclass[10pt,a4paper]{article}
\usepackage[utf8]{inputenc}
\usepackage{amsmath}
\usepackage{amsfonts}
\usepackage{amssymb}
\usepackage{graphicx}
\usepackage{fullpage}

\newcommand{\mc}[1]{\mathcal{#1}} 
\newcommand{\mtxt}[1]{\quad\text{#1}\quad} 
\newcommand{\set}[2]{\left\lbrace #1 \;|\; #2 \right\rbrace} 
\newcommand{\collect}[1]{\lbrace#1\rbrace}
\newcommand{\norm}[1]{\|#1\|}
\newcommand{\euc}[1]{\mathbb{R}^{#1}} 
\newcommand{\sph}[1]{\mathbb{S}^{#1}} 
\newcommand{\hph}{\mathbb{H}^{2}} 
\newcommand{\vol}[1]{\mathrm{vol}\left(#1\right)} 

\renewcommand{\v}[1]{\ensuremath{\mathbf{#1}}} 
 
\newcommand{\abs}[1]{\left| #1 \right|} 
\renewcommand{\d}[2]{\frac{d #1}{d #2}} 
\newcommand{\pd}[2]{\frac{\partial #1}{\partial #2}} 
\newcommand{\pdd}[2]{\frac{\partial^2 #1}{\partial #2^2}} 
\newcommand{\ket}[1]{\left| #1 \right>} 
\newcommand{\braket}[2]{\left< #1 \vphantom{#2} \right|
 \left. #2 \vphantom{#1} \right>} 
\newcommand{\matrixel}[3]{\left< #1 \vphantom{#2#3} \right|
 #2 \left| #3 \vphantom{#1#2} \right>} 
\let\baraccent=\= 
\renewcommand{\=}[1]{\stackrel{#1}{=}} 

\title{Exact Renormalization Group for Point Interactions}
\author{Osman Teoman Turgut and Cem Eröncel\footnote{Present Address: Syracuse University, Department of Physics, Syracuse, NY 13244} \\
\small Bogazici University, Department of Physics, 34342 Bebek, Istanbul}
\begin{document}
\maketitle
\begin{abstract}
Renormalization is one of the deepest ideas in physics, yet its exact implementation in any interesting problem is usually very hard. In the present work, following the approach by Glazek and Maslowski in the flat space, we will study the exact renormalization of the same  problem in a nontrivial geometric setting, namely in the two dimensional hyperbolic space. Delta function potential is an asymptotically free quantum mechanical problem which makes it resemble nonabelian gauge
theories, yet it can be treated exactly in this nontrivial geometry. 
\end{abstract}
\section{Introduction}
Most problems of deep significance in the world of interacting many-particles are formulated by singular theories. Typically, these are plagued by divergences, which reflects our ignorance of the physics beyond the scales defined by our original theory. 

A deep insight into this behavior came from Wilson  \cite{Wil0, Wil1, Wil2, Wil3, Wil4, WilKog_1974}. He argued that the physics beyond the scales of interest should be incorporated into lower energies by some effective interactions. As we will show in Section \ref{hamrenom}, for a system defined by a Hamiltonian, if one calculates the form of the effective Hamiltonian at some energy scale $\Lambda$ specifying the cutoff, the result will be
\begin{align*}
H_{\textrm{eff}}^{\Lambda}=\mathcal{P}H\mathcal{P}+X_{\Lambda},
\end{align*}
where $\mathcal{P}H\mathcal{P}$ is the projection of the Hamiltonian on the subspace where momentum eigenvalues are bounded by the cutoff $\Lambda$, and the operator $X_{\Lambda}$ depends on higher degrees of freedom. Since the cutoff is totally arbitrary, the effective Hamiltonian should not depend on it. The essence of Wilsonian Renormalization Group, or Exact Renormalization Group (ERG), is modifying the parameters of the theory without altering the energy eigenvalues, so that the effective Hamiltonian becomes cutoff-independent. This is done as follows: We start by specifying the system at some high energy scale $\Lambda$, called the \emph{bare scale}. Then we introduce another scale $\lambda$, called the \emph{effective scale}, such that $1\ll\lambda\ll\Lambda$. The ERG procedure consists of integrating out degrees of freedom between these two scales. This integrating out procedure is not performed in a single step. In each step one integrates over an infinitesimal momentum shell. This transformation, which is called a \emph{Renormalization Group (RG) transformation} creates a trajectory, called \emph{RG trajectory} in the space of theories, or in this particular case in the space of Hamiltonians. The Hamiltonians at different scales are related by the requirement that the eigenvalues do not change as one changes the scale. In other words the RG trajectory is determined by the condition that all the Hamiltonians on this trajectory give the same set of eigenvalues as the unrenormalized theory.

There is no systematic non-perturbative approach to implement this idea yet to this date, nevertheless many interesting problems can be solved by means of some approximation method. The literature in this direction is immense, and we don't feel competent enough to cite all the relevant works. We will mention just a few things related to the present work. A perturbative approach to the renormalization group for effective Hamiltonians in light-front field theory is given in \cite{GlaWil94}. Another renormalization procedure for light-front Hamiltonians is called \emph{Similarity Renormalization Group}, where the bare Hamiltonian with an arbitrary large, but finite cutoff, is transformed by a similarity transformation which makes the Hamiltonian \emph{band diagonal} \cite{GlaWil93, GlaWil98}. A pedagogical treatment can be found in \cite{Gla08}.

One of the main challenges is to understand renormalization in a system where the interactions lead to the appearance of bound states. Indeed quantum chromodynamics is the main example we have in mind, where the theory is formulated in terms of physically unobservable variables, in ordinary energy scales, as a result of interactions only their bound states become physical particles. Since one is interested in understanding the formation of these bound states and calculating the resulting masses, in principle, it is most natural to work with the Hamiltonian directly. Of course this is a very hard problem. As a result it is valuable and interesting to learn more about renormalization and its non-perturbative aspects even in very simple systems using the Hamiltonian formalism. This has been done by Glazek and Maslowski for the Dirac-delta function in two dimensions \cite{Glazek}. In the present work, we will consider the same problem on a nontrivial manifold, two dimensional hyperbolic space. This is interesting because the gauge theory problem also has a nontrivial geometry when it is formulated on the space of connections modulo gauge equivalent configurations \cite{mitter1981}. Hence, it is a nice exercise to see that type of complications may arise when the underlying geometry is nontrivial. 

We shall start by reviewing point interactions on the Euclidean Plane and answer why this problem requires renormalization. following \cite{Glazek}, we will review the renormalization of point interactions in the Euclidean plane using the Wilsonian RG scheme and derive the flow equation. As an addendum to Glazek and Maslowski, we also investigate the range of renormalizability using the Banach Contraction principle. In the next section we shall analyze the same problem on the hyperbolic plane and show that the flow equation has the same form. Finally we will speak about a puzzle where this procedure fails at a technical level, if one study the same problem on a compact manifold, namely two-dimensional sphere, $\mathbb{S}^{2}$.
\section{Exact Renormalization Group on the Euclidean Plane}
\subsection{Formulation of the Problem}
The Schrödinger equation for the simplest type of point interaction on a $D$-dimensional Euclidean space $\euc{D}$ is given by, in units $\hbar=1$ and $2m=1$, as
\begin{align}\label{schd1_2}
\left(-\Delta_{\euc{D}} - g\delta^D(\v{x})\right)\psi(\v{x})=E\psi(\v{x}),
\end{align}
where $\Delta_{\euc{D}}$ is the Laplacian operator on $\euc{D}$ and $g$ is real, positive parameter which determines the strength of the point interaction. If we parametrize the bound state energy by $E=-\nu^2$, then the Schrödinger equation for the bound state of the system becomes
\begin{align}\label{schd2}
\left(-\Delta_{\euc{D}} - g\delta^D(\v{x})\right)\phi(\v{x})=-\nu^2\phi(\v{x}),
\end{align}
where $\phi(\v{x})$ is the bound state wavefunction. This expression can be expressed in momentum space as
\begin{align}\label{schd3}
(p^2+\nu^2)\tilde{\phi}(p,\omega)=\frac{g}{(2\pi)^D}\int_{\sph{D-1}}d\omega\int_{0}^{\infty}dp'\,p'^{D-1}\tilde{\phi}(p',\omega),
\end{align}
where $\tilde{\phi}(p,\omega)$ is the Fourier Transform of $\phi(\v{x})$. Note that we also switched to spherical coordinates in momentum space, where $p$ and $\omega$ denote the radial and angular coordinates respectively. From \eqref{schd3}, $g^{-1}$ can be solved as
\begin{align}\label{schd5}
\frac{1}{g}=\frac{\vol{\sph{D-1}}}{(2\pi)^D}\int_{0}^{\infty}dp'\,\frac{p'^{D-1}}{p'^2+\nu^2},
\end{align}
where $\vol{\sph{D-1}}$ denotes the volume of the unit sphere in $D-1$ dimensions. It is easy to see that the integral diverges for $D\geq 2$, so regularization and renormalization are needed to obtain physical results. 

Renormalization of point interactions has been studied by many authors; in position space \cite{GosTar_1991, Han_1993, MeaGod_1991}, and in momentum space \cite{AdhGho_1997, AraTomAdhFre_1997, Cav_2000, MitGupRoy_1998, Nye_2000}. The renormalization group equations were derived in \cite{AdhFre_1995, AdhGho_1997}. Instead of the conventional approach, we shall perform the renormalization using the Exact Renormalization Group (ERG) method.
\subsection{Renormalization of Hamiltonians}\label{hamrenom}
In this section we perform the renormalization of a point interaction on the Euclidean plane, $\euc{2}$. This part will be mainly a review of the lecture notes given by G\l{}azek and Maslowski \cite{Glazek} and we include it for the sake of completeness. However our approach will be slightly different and as an addendum we also investigate the range of renormalizability using the Banach Contraction Principle. 

Before any kind of regularization or renormalization, the Schrödinger equation for the bound state can be written as
\begin{align}\label{reham01}
H\ket{\phi}=-\nu^2\ket{\phi}.
\end{align}
We want to calculate the effective Hamiltonian $H^{\Lambda}_{\textrm{eff}}$ at some energy scale $\Lambda$, where $\Lambda \gg 1$. This is done by integrating out degrees of freedom above $\Lambda$. To this end we introduce the operators $\mc{P}$ and $\mc{Q}$ which are projections to the subspaces, where the momentum eigenvalue takes the values $0\leq p \leq \Lambda$ and $p > \Lambda$ respectively. Let us also define $\ket{\phi}_{\mc{P}}\equiv \mc{P}\ket{\phi}$ and $\ket{\phi}_{\mc{Q}}\equiv \mc{Q}\ket{\phi}$. By using $\mc{P}+\mc{Q}=I$ and $\mc{P}\mc{Q}=0$ one can split \eqref{reham01} as
\begin{align}
&\mc{P}H\mc{P}\ket{\phi}_{\mc{P}}+\mc{P}H\mc{Q}\ket{\phi}_{\mc{Q}}=-\nu^2\ket{\phi}_{\mc{P}} \label{reham02}\\
&\mc{Q}H\mc{P}\ket{\phi}_{\mc{P}}+\mc{Q}H\mc{Q}\ket{\phi}_{\mc{Q}}=-\nu^2\ket{\phi}_{\mc{Q}} \label{reham03}.
\end{align}
From \eqref{reham03} we find
\begin{align}
\ket{\phi}_{\mc{Q}}=(-\nu^2-\mc{Q}H\mc{Q})^{-1}\mc{Q}H\mc{P}\ket{\phi}_{\mc{P}}.
\end{align}
If we substitute this result back into \eqref{reham02} we get
\begin{align}\label{reham04}
\left(\mc{P}H\mc{P}+\mc{P}H\mc{Q}(-\nu^2-\mc{Q}H\mc{Q})^{-1}\mc{Q}H\mc{P}\right)\ket{\phi}_{\mc{P}}=-\nu^2\ket{\phi}_{\mc{P}}
\end{align}
and this implies that the effective Hamiltonian at the scale $\Lambda$ is given by
\begin{align}\label{reham05}
H^{\Lambda}_{\textrm{eff}}=\mc{P}H\mc{P}+\mc{P}H\mc{Q}(-\nu^2-\mc{Q}H\mc{Q})^{-1}\mc{Q}H\mc{P}\equiv \mc{P}H\mc{P} + X_{\Lambda}.
\end{align}
We note that, although we are working at the scale $\Lambda$, the effective Hamiltonian contains the $X_{\Lambda}$ term and this term, which is called \emph{counterterm}, depends on the higher degrees of freedom. And as we shall see now, we will use this counterterm in order to define the effective coupling constant at the scale $\Lambda$. Let us write the Hamiltonian as $H=H_0+V$ where $H_0$ is the free Hamiltonian and $V$ denotes the point interaction, i.e. $\matrixel{\v{x}}{V}{\phi}=-g \delta^{2}(\v{x})\phi(\v{x})$. \eqref{reham04} can be written in momentum space as
\begin{align}
(p^2+\nu^2)\tilde{\phi}_{\mc{P}}(\v{p})+\int_{\euc{D}}d^D\v{p}'\matrixel{\v{p}}{\mc{P}V\mc{P}}{\v{p}'}\tilde{\phi}_{\mc{P}}(\v{p}')+\int_{\euc{D}}d^D\v{p}'\,\matrixel{\v{p}}{X_{\Lambda}}{\v{p}'}\tilde{\phi}_{\mc{P}}(\v{p}')=0
\end{align}
where $\tilde{\phi}_{\mc{P}}(\v{p})=\braket{\v{p}}{\tilde{\phi}_{\mc{P}}}$. By defining $x_{\Lambda}(\v{p},\v{p}')\equiv(2\pi)^2 \matrixel{\v{p}}{X_{\Lambda}}{\v{p}'}$ and using
\begin{align}\label{deltmom}
\matrixel{\v{p}}{V}{\v{p'}}=\int_{\euc{4}}d^2\v{x}\,d^2\v{x}'\,\braket{\v{p}}{\v{x}}\matrixel{\v{x}}{V}{\v{x}'}\braket{\v{x}'}{\v{p}'} = -\frac{g}{(2\pi)^2},
\end{align}
we get
\begin{align}
(p^2+\nu^2)\tilde{\phi}_{\mc{P}}(\v{p})-\frac{1}{(2\pi)^2}\int_{\euc{2}}d^2\v{p}'\Theta_{\Lambda}(\v{p})\left(g-x_{\Lambda}(\v{p},\v{p}')\right)\tilde{\phi}_{\mc{P}}(\v{p}')=0,
\end{align}
where $\Theta_{\Lambda}(\v{p})$ is the step function. We see that the $g-x_{\Lambda}(\v{p},\v{p}')$ term plays the role of the effective coupling constant. From now on we denote it by $g_{\Lambda}(\v{p},\v{p}')$. The counterterm $x_{\Lambda}(\v{p},\v{p}')$ acts like a correction to the initial theory and by using it we have defined the renormalized coupling constant $g_{\Lambda}(\v{p},\v{p}')$ at the scale $\Lambda$.
\subsection{Applying the ERG Procedure}
Now we are in a position to perform the ERG analysis of our theory. Since the original problem is rotationally symmetric, we want to keep rotational symmetry intact. Therefore we assume that the renormalized coupling constant $g_{\Lambda}$ does not depend on $\omega$. At the bare scale $\Lambda$ we can write the following equation:
\begin{align}\label{erg1}
(p^2+\nu^2)\tilde{\phi}(p,\omega)=\frac{\Theta_{\Lambda}(p)}{(2\pi)^2}\int_0^{\Lambda}dp'\,p'g_{\Lambda}(p,p')\vartheta(p')
\end{align}
where
\begin{align}\label{erg2}
\vartheta(p)\equiv\int_{\sph{1}}d\omega\,\tilde{\phi}(p,\omega).
\end{align}
We remark that we have switched to the unprojected wavefunction $\tilde{\phi}(p,\omega)$ and compensate this change by putting the step function $\Theta_{\Lambda}(p)$ in front of the integral, which ensures that \eqref{erg1} is valid for $p\leq\Lambda$. Following the ERG procedure, we write the analog of \eqref{erg1} at the infinitesimally lower scale $\Lambda-d\Lambda$.
\begin{align}\label{erg3}
(p^2+\nu^2)\tilde{\phi}(p,\omega)=\frac{\Theta_{\Lambda-d\Lambda}(p)}{(2\pi)^2}\int_0^{\Lambda-d\Lambda}dp'\,p'g_{\Lambda-d\Lambda}(p,p')\vartheta(p').
\end{align}
We can rewrite \eqref{erg1} as
\begin{align}\label{erg4}
(p^2+\nu^2)\tilde{\phi}(p,\omega)=\frac{\Theta_{\Lambda}(p)}{(2\pi)^2}\left(\int_0^{\Lambda-d\Lambda}dp'\,p'g_{\Lambda}(p,p')\vartheta(p')+d\Lambda\, \Lambda\, g_{\Lambda}(p,\Lambda)\vartheta(\Lambda)\right).
\end{align}
For $p=\Lambda$ this will give us
\begin{align}\label{erg5}
(\Lambda^2+\nu^2)\tilde{\phi}(\Lambda,\omega)=\frac{1}{(2\pi)^2}\left(\int_0^{\Lambda-d\Lambda}dp'\,p'g_{\Lambda}(\Lambda,p')\vartheta(p')+d\Lambda\, \Lambda\, g_{\Lambda}(\Lambda,\Lambda)\vartheta(\Lambda)\right),
\end{align}
and from this we can read of $\tilde{\phi}(\Lambda,\omega)$ as
\begin{align}\label{erg6}
\tilde{\phi}(\Lambda,\omega)=\frac{1}{(2\pi)^2(\Lambda^2+\nu^2)}\int_0^{\Lambda-d\Lambda}dp'\,p'g_{\Lambda}(\Lambda,p')\vartheta(p')
\end{align}
where we have ignored the term which is proportional to $d\Lambda$. If we substitute this result into \eqref{erg2} and perform the $\omega$ integral we find
\begin{align}
\vartheta(\Lambda)=\frac{1}{(2\pi)(\Lambda^2+\nu^2)}\int_0^{\Lambda-d\Lambda}dp'\,p'g_{\Lambda}(\Lambda,p')\vartheta(p').
\end{align}
Finally we put this result into \eqref{erg4} to obtain
\begin{align}
(p^2+\nu^2)\tilde{\phi}(p,\omega)=\frac{\Theta_{\Lambda}(p)}{(2\pi)^2}\int_0^{\Lambda-d\Lambda}\!\!\!&dp'\,p'\left(g_{\Lambda}(p,p')+\frac{d\Lambda\,\Lambda}{2\pi(\Lambda^2+\nu^2)}g_{\Lambda}(p,\Lambda)g_{\Lambda}(\Lambda,p')\right)\vartheta(p')
\end{align}
Clearly, we can replace $\Theta_{\Lambda}(p)$ by $\Theta_{\Lambda-d\Lambda}(p)$ and write
\begin{align}
(p^2+\nu^2)\tilde{\phi}(p,\omega)=\frac{\Theta_{\Lambda-d\Lambda}(p)}{(2\pi)^2}\int_0^{\Lambda-d\Lambda}\!\!\!&dp'\,p'\left(g_{\Lambda}(p,p')+\frac{d\Lambda\,\Lambda}{2\pi(\Lambda^2+\nu^2)}g_{\Lambda}(p,\Lambda)g_{\Lambda}(\Lambda,p')\right)\vartheta(p')
\end{align}
Now comparing this equation with \eqref{erg3} gives us an equation for the coupling constant
\begin{align}
g_{\Lambda-d\Lambda}(p,p')=g_{\Lambda}(p,p')+\frac{d\Lambda \,\Lambda}{2\pi(\Lambda^2+\nu^2)}g_{\Lambda}(p,\Lambda)g_{\Lambda}(\Lambda,p'),
\end{align}
which can be put into differential form as
\begin{align}
-\d{g_{\Lambda}(p,p')}{\Lambda}=\frac{\Lambda}{2\pi(\Lambda^2+\nu^2)}g_{\Lambda}(p,\Lambda)g_{\Lambda}(\Lambda,p').
\end{align}
This equation determines the RG trajectory of the coupling constant. To find the effective coupling at the effective scale $\lambda$ we integrate this from $\lambda$ to $\Lambda$ and find
\begin{align}
g_{\lambda}(p,p')=g_{\Lambda}(p,p')+\frac{1}{2\pi}\int_{\lambda}^{\Lambda}ds\,\frac{s}{s^2+\nu^2}g_{s}(p,s)g_{s}(s,p')
\end{align}
or
\begin{align}\label{erg7}
g_{\lambda}(p,p')=g-x_{\Lambda}(p,p')+\frac{1}{2\pi}\int_{\lambda}^{\Lambda}ds\,\frac{s}{s^2+\nu^2}g_{s}(p,s)g_{s}(s,p')
\end{align}
Although this is an ordinary differential equation with three variables and we have one initial condition, there is also the requirement that $g_{\lambda}(p,p')$ should not depend on $\Lambda$ when we take the $\Lambda\rightarrow \infty$ limit. This can be satisfied by the appropriate choice of the counterterm $x_{\Lambda}(p,p')$. We try an iteration procedure to obtain a solution.

At the first order we choose
\begin{align}
g_{\lambda}^{(1)}(p,p')=g\mtxt{so that}x_{\Lambda}^{(1)}=0.
\end{align}
After substituting these choices to \eqref{erg7} we get
\begin{align}
g_{\lambda}^{(2)}(p,p')=g-x_{\Lambda}^{(2)}(p,p')+\frac{g^2}{2\pi}\int_{\lambda}^{\Lambda}ds\,\frac{s}{s^2+\nu^2}.
\end{align}
The integral diverges in the $\Lambda\rightarrow \infty$ limit, therefore we choose the counterterm as
\begin{align}
x_{\Lambda}^{(2)}(p,p')=\frac{g^2}{2\pi}\int_{\lambda_0}^{\Lambda}ds\,\frac{s}{s^2+\nu^2},
\end{align}
where $\lambda_0$ is an another energy scale chosen such that $1\ll \lambda_0 < \lambda \ll \Lambda$. Now the effective coupling at the second order is finite and given by
\begin{align}
g_{\lambda}^{(2)}(p,p')=g-\frac{g^2}{2\pi}\int_{\lambda_0}^{\lambda}ds\,\frac{s}{s^2+\nu^2}.
\end{align}
We note that it is independent of $p$ and $p'$. If we repeat this procedure, then by induction it is straightforward to see that $g_{\lambda}^{(n)}$ and $x_{\Lambda}^{(n)}$ are independent of $p$ and $p'$ for all $n$. At the order $n+1$, the effective coupling becomes
\begin{align}\label{coup_n}
g_{\lambda}^{(n+1)}=g-x_{\Lambda}^{(n+1)}+\frac{1}{2\pi}\int_{\lambda}^{\Lambda}ds\,\frac{s}{s^2+\nu^2}\left(g_s^{(n)}\right)^2.
\end{align}
We choose the counterterm as
\begin{align}
x_{\Lambda}^{(n+1)}=\frac{1}{2\pi}\int_{\lambda_0}^{\Lambda}ds\,\frac{s}{s^2+\nu^2}\left(g_s^{(n)}\right)^2,
\end{align}
hence we find
\begin{align}
g_{\lambda}^{(n+1)}=g-\frac{1}{2\pi}\int_{\lambda_0}^{\lambda}ds\,\frac{s}{s^2+\nu^2}\left(g_s^{(n)}\right)^2.
\end{align}
It is not trivial to conclude that this iteration process has a limit. We shall deal with this later in this section and for now we assume that $g$ and $\lambda_0$ are chosen such that the sequence $\lbrace g_{\lambda}^{(n)}\rbrace_{n=1}^{\infty}$ has a limit given by
\begin{align}
\lim_{n\rightarrow\infty}g_{\lambda}^{(n+1)}=g_{\lambda}.
\end{align}
After taking the limit we can write for the effective coupling
\begin{align}
g_{\lambda}=g-\frac{1}{2\pi}\int_{\lambda_0}^{\lambda}ds\,\frac{s}{s^2+\nu^2}g_s^2,
\end{align}
which immediately implies $g=g_{\lambda_0}$. This equation can be put into the following form
\begin{align}
\int_{\lambda_0}^{\lambda}ds\,\d{g_s}{s}=-\frac{1}{2\pi}\int_{\lambda_0}^{\lambda}ds\,\frac{s}{s^2+\nu^2}g_{s}^2,
\end{align}
which implies
\begin{align}
\frac{dg_s}{g_s^2}=-\frac{1}{2\pi}\frac{s\,ds}{s^2+\nu^2}.
\end{align}
After integrating this equation from $\lambda_0$ to $\lambda$ and solving for $g_{\lambda}$, we obtain the final answer.
\begin{align}\label{r_final}
g_{\lambda}=\frac{g_{\lambda_0}}{1+\frac{g_{\lambda_0}}{4\pi}\log\left(\frac{\lambda^2+\nu^2}{\lambda_0^2+\nu^2}\right)}
\end{align}
This result is in agreement with the one given in \cite{AdhFre_1995}. We also note that as $\lambda\rightarrow \infty$, $g_{\lambda}\rightarrow 0$, so the theory is asymptotically free.

\subsection{Estimating the Range of Renormalizability}\label{cont_r}
Now we shall try to investigate under which conditions the sequence $\lbrace g_{\lambda}^{(n)}\rbrace_{n=1}^{\infty}$ has a limit. However we don't have a closed form expression for $g_{\lambda}^{(n)}$, \eqref{coup_n} tells us that $g_{\lambda}^{(n)}$ depends on $g_{\lambda}^{(n-1)}$, hence we could not obtain a rigorous result for the convergence radius of the series $\collect{g_{\lambda}^{(n)}}_{n=1}^{\infty}$ in this way. An alternative way is to investigate under which circumstances does the integral equation given by
\begin{align}\label{ban1}
g_{\lambda}=g_{\lambda_0}-\frac{1}{2\pi}\int_{\lambda_0}^{\lambda}ds\,\frac{s}{s^2+\nu^2}\,g_s^2
\end{align}
has a unique solution. This can be done by using the theory of ordinary differential equations. We begin by defining a compact interval $I=[\lambda_0,\tilde{\lambda}]\subset \euc{}$ where $1\ll\lambda_0<\lambda<\tilde{\lambda}\ll\Lambda$. Let $C(I)$ denote the space of continuous functions on $I$. It becomes a vector space if the vector space operations are defined pointwise. Moreover it is well known that it is a Banach Space if we define a norm on $C(I)$ by
\begin{align}
\norm{\mathfrak{g}}=\sup_{s\in I}\abs{g_s}.
\end{align}
We note that we use $\mathfrak{g}$, $\mathfrak{h}$ as elements of $C(I)$ to avoid confusion with the unrenormalized coupling constant $g$, that is, we made the definition $\mathfrak{g}(s)\equiv g_s$. Now we introduce a map $T:C(I)\rightarrow C(I)$ defined by
\begin{align}\label{ban2}
T(\mathfrak{g})(\lambda)=g_{\lambda_0}-\frac{1}{2\pi}\int_{\lambda_0}^{\lambda}ds\,\frac{s}{s^2+\nu^2}\,g_s^2
\end{align}
Then \eqref{ban1} can be expressed as $\mathfrak{g}=T\mathfrak{g}$, in other words the solution of \eqref{ban1} is also a \emph{fixed point} of $T$. The existence and uniqueness of a solution to $\mathfrak{g}=T\mathfrak{g}$ can be proved using \emph{Contraction Principle} which can be stated as follows: Let $D$ be a nonempty closed subset of a Banach Space $\mc{B}$. If a map $T:D\rightarrow \mc{B}$ is a contraction and maps $D$ into itself, i.e. $T(D)\subseteq D$, then $T$ has an exactly one fixed point $\overline{\mathfrak{g}}$ which is in $D$ \cite{ODE}. $T$ is a contraction means that there exist a positive constant $\theta<1$ such that
\begin{align}
\norm{T(\mathfrak{g})-T(\mathfrak{h})}\leq \theta \norm{\mathfrak{g}-\mathfrak{h}},\mtxt{for}\mathfrak{g},\mathfrak{h}\in D.
\end{align}
If $T$ is a contraction, then the sequence $\collect{\mathfrak{g}^{(n)}}_{n=1}^{\infty}$ defined by
\begin{align}
\mathfrak{g}^{(n)}=T(\mathfrak{g}^{(n-1)})\mtxt{with}\mathfrak{g}^{(1)}=T(\mathfrak{g}_0),
\end{align}
where $\mathfrak{g}_0$ is an arbitrary element of $D$, converges to the fixed point $\overline{\mathfrak{g}}$, that is
\begin{align}
\lim_{n\rightarrow \infty}\norm{\mathfrak{g}^{(n)}-\overline{\mathfrak{g}}}=0.
\end{align}
Therefore to estimate the range of renormalizability of our theory, we need to estimate under which cases the map $T$ defined as in \eqref{ban2} is a contraction. First of all we need a closed subset of $C(I)$. From \eqref{ban1} we can conclude that if $\mathfrak{g}$ is a solution, then it should be monotone decreasing on $I=[\lambda_0,\tilde{\lambda}]$. Thus it is natural to choose our closed subset as
\begin{align}
D=\set{\mathfrak{g}\in C(A)}{\norm{\mathfrak{g}}\leq g_{\lambda_0}}.
\end{align}
Then
\begin{align}
\norm{T(\mathfrak{g})}=\sup_{\lambda\in I}\abs{T(\mathfrak{g})(\lambda)}=\sup_{\lambda\in I}\abs{g_{\lambda_0}-\frac{1}{2\pi}\int_{\lambda_0}^{\lambda}ds\,\frac{s}{s^2+\nu^2}\,g_s^2}=g_{\lambda_0},
\end{align}
thus $T(D)\subseteq D$. So it remains to show that $T$ is a contraction. Let $\mathfrak{g},\mathfrak{h}\in D$. Then we have the estimate
\begin{align}
\abs{T(\mathfrak{g})-T(\mathfrak{h})}&=\frac{1}{2\pi}\int_{\lambda_0}^{\lambda}ds\,\frac{s}{s^2+\nu^2}\left((h_s)^2-(g_s)^2\right)\nonumber \\
&\leq\frac{1}{2\pi}\sup_{s\in [\lambda_0,\lambda]}\abs{(h_s)^2-(g_s)^2}\int_{\lambda_0}^{\lambda}ds\,\frac{s}{s^2+\nu^2}\nonumber \\
&\leq\frac{1}{2\pi}\sup_{s\in [\lambda_0,\lambda]}\abs{(h_s)^2-(g_s)^2}\int_{\lambda_0}^{\lambda}ds\,\frac{1}{s}\nonumber \\
&=\frac{1}{2\pi}\sup_{s\in [\lambda_0,\lambda]}\abs{(h_s+g_s)(h_s-g_s)}\log\left(\frac{\lambda}{\lambda_0}\right)
\end{align}
By taking the supremum of both sides we find
\begin{align}
\norm{T(\mathfrak{g})-T(\mathfrak{h})}&\leq
\frac{1}{2\pi}\sup_{s\in I}\abs{(h_s+g_s)(h_s-g_s)}\log\left(\frac{\tilde{\lambda}}{\lambda_0}\right)\nonumber \\
&=\frac{1}{2\pi}\norm{\mathfrak{g}+\mathfrak{h}}\norm{\mathfrak{g}-\mathfrak{h}}\log\left(\frac{\tilde{\lambda}}{\lambda_0}\right)\nonumber \\
&\leq\frac{1}{2\pi}(2g_{\lambda_0})\norm{\mathfrak{g}-\mathfrak{h}}\log\left(\frac{\tilde{\lambda}}{\lambda_0}\right).
\end{align}
This tells us that $T$ is a contraction if
\begin{align}\label{ban4}
\frac{g_{\lambda_0}}{\pi}\log\left(\frac{\tilde{\lambda}}{\lambda_0}\right)<1.
\end{align}
If we interpret the interval $I=[\lambda_0,\tilde{\lambda}]$ as the range of renormalizability, then from \eqref{ban4} we can see that it is directly related to the coupling at the energy scale $\lambda_0$. For a small coupling $g_{\lambda_0}\ll 1$, we can shift up $\tilde{\lambda}$ considerably without breaking the contraction property of $T$, however for couplings $g_{\lambda_0}\sim 1$, the range is quite small or we may not even prove the existence of a solution by this approach. 
\subsection{Bound State Solution}
We can check that with the coupling constant given as in \eqref{r_final} we get a finite answer for the bound state energy. For this we plug \eqref{r_final} into \eqref{erg1} to find
\begin{align}\label{ergb1}
(p^2+\nu^2)\tilde{\phi}(p,\omega)=\frac{\Theta_{\Lambda}(p)}{(2\pi)^2}\frac{g_{\lambda_0}}{1+\frac{g_{\lambda_0}}{4\pi}\log\left(\frac{\Lambda^2+\nu^2}{\lambda^2_0+\nu^2}\right)}\int_{0}^{\Lambda}dp'\,p'\int_{\sph{1}}d\omega'\,\tilde{\phi}(p',\omega').
\end{align}
By defining
\begin{align}\label{ergb2}
\mc{N}=\int_{0}^{\Lambda}dp'\,p'\int_{\sph{1}}d\omega'\,\tilde{\phi}(p',\omega'),
\end{align}
we obtain
\begin{align}\label{ergb3}
\tilde{\phi}(p,\omega)=\frac{\Theta_{\Lambda}(p)}{(2\pi)^2}\frac{g_{\lambda_0}}{1+\frac{g_{\lambda_0}}{2\pi}\log\left(\frac{\Lambda}{\lambda_0}\right)}\frac{\mc{N}}{p^2+\nu^2}.
\end{align}
Substituting this result into \eqref{ergb2} and diving both sides by $\mc{N}$ give us
\begin{align}
1=\frac{1}{4\pi}\frac{g_{\lambda_0}}{1+\frac{g_{\lambda_0}}{4\pi}\log\left(\frac{\Lambda^2+\nu^2}{\lambda_0^2+\nu^2}\right)}\log\left(\frac{\Lambda^2+\nu^2}{\nu^2}\right).
\end{align}
From this equation we can solve for $\nu^2$ and in the $\Lambda\rightarrow\infty$ limit and find
\begin{align}
E_b=\lim_{\Lambda\rightarrow\infty}-\nu^2=-\lambda_0^2\frac{e^{-4\pi/g_{\lambda_0}}}{1-e^{-4\pi/g_{\lambda_0}}}
\end{align}
which is finite.
\section{Exact Renormalization Group on the Hyperbolic Plane}
We will begin this section by constructing the spectral representation of the Laplacian on the hyperbolic plane $\hph$. By using this construction we shall perform the ERG analysis of a point interaction on the hyperbolic plane. 
\subsection{The Geometry and Spectra of the Hyperbolic Plane}
We shall do the construction by using ideas given in \cite{His_1994} and \cite{Audrey}. There are various models for the hyperbolic plane. We will use the \emph{upper half-plane} model, where $\hph$ is realized as the set
\begin{align}
\hph = \set{z=(x,y)}{x\in \mathbb{R}\,, y\in [0,\infty)},
\end{align}
with the Riemannian metric $g_{\hph}$ given by
\begin{align}
g_{\hph}=\frac{R^2}{y^2}\begin{pmatrix}
1 & 0 \\ 0 & 1
\end{pmatrix},
\end{align}
where $-R^{-2}$ is the constant sectional curvature. The Riemannian volume element is given by 
\begin{align}
dV_{\hph}=\sqrt{\det g_{\hph}}\,dx\wedge dy=\frac{dx\,dy}{y^2/R^2},
\end{align}
and the Laplacian is
\begin{align}
\Delta_{\hph}=\frac{y^2}{R^2}\left(\pdd{}{x}+\pdd{}{y}\right).
\end{align}
The eigenfunctions can be found by solving the closed eigenvalue problem on $L^2(\hph,dV_{\hph})$ expressed by
\begin{align}\label{spec1}
(\Delta_{\hph}+\lambda)f(z)=0,
\end{align}
where $\lambda\in \euc{}$. For notational simplicity, let us define $\tilde{\Delta}_{\hph}\equiv R^2\Delta_{\hph}$ and $\tilde{\lambda}\equiv R^2\lambda$. Then \eqref{spec1} will be equivalent to
\begin{align}\label{spec1_2}
(\tilde{\Delta}_{\hph}+\tilde{\lambda})f(z)=0,
\end{align}
  Since $\tilde{\Delta}_{\hph}$ is separable in $(x,y)$ coordinates we can use separation of variables. So we choose $f(z)=v(x)w(y)$ and put this into \eqref{spec1_2} to obtain
\begin{align}
\pdd{v}{x}\frac{1}{v(x)}+\pdd{w}{y}\frac{1}{w(y)}+\frac{\tilde{\lambda}}{y^2}=0.
\end{align}
This implies that there is a constant $\xi^2$ such that
\begin{align}\label{spec2}
\pdd{v}{x}\frac{1}{v(x)}=-\xi^2\mtxt{and}\pdd{w}{y}\frac{1}{w(y)}+\frac{\tilde{\lambda}}{y^2}=\xi^2.
\end{align}
The $x$-part can be solved easily as $v(x)=e^{i\xi x}$. To solve the $y$-part we introduce a new function by $u(y)\equiv y^{-1/2}w(y)$. After substituting this into the $y$-part of \eqref{spec2} and making some rearrangements we get
\begin{align}\label{spec3}
y^2\pdd{u}{y}+y\pd{u}{y}-\left(y^2\xi^2+\frac{1}{4}-\tilde{\lambda}\right)u(y)=0
\end{align}
The eigenvalues of the Laplacian on $\hph$ starts with $\tilde{\lambda}_0=\frac{1}{4}$ \cite{Bort}. Therefore $\frac{1}{4}-\tilde{\lambda}\leq 0$ so we introduce a new variable $\tau\in [0,\infty)$ such that $\frac{1}{4}-\tilde{\lambda}=(i\tau)^2$. Then \eqref{spec3} becomes
\begin{align}\label{spec4}
y^2\pdd{u}{y}+y\pd{u}{y}-\left[(y\xi)^2+(i\tau)^2\right]u(y)=0.
\end{align}
There are two linearly independent solutions which are the modified Bessel Functions $I_{i\tau}(\abs{y\xi})$ and $K_{i\tau}(\abs{y\xi})$. Since $I_{i\tau}(\abs{y\xi})$ is singular at infinity, we exclude it from our solution space. Moreover $K_{i\tau}(\abs{y\xi})$ has a singularity at $\xi=0$ given by \cite{Audrey}
\begin{align}
K_{i\tau}(\abs{y\xi})\sim 2^{i\tau-1}\Gamma(i\tau)\abs{y\xi}^{-i\tau}+2^{-i\tau-1}\Gamma(-i\tau)\abs{y\xi}^{i\tau}\quad \textrm{as }\xi\rightarrow 0^+.
\end{align}
So we choose $u(y)=\abs{\xi}^{i\tau}K_{i\tau}(\abs{y\xi})$ as a solution to \eqref{spec4} and write the eigenfunctions of $\Delta_{\hph}$ as
\begin{align}\label{heig}
E_0(z;\tau,\xi)=\frac{1}{\sqrt{2\pi}}e^{i\xi x}\sqrt{y}\abs{\xi}^{i\tau}K_{i\tau}(\abs{y\xi})
\end{align}
where we have put an extra $(2\pi)^{-1/2}$ to simplify our construction of the spectral representation. We note that this does not alter the spectrum of the eigenvalues. 

To obtain the spectral representation of $\Delta_{\hph}$ we introduce the following transform,
\begin{align}
&(\mc{K}\psi)(\tau,\xi)\equiv \tilde{\psi}(\tau,\xi)=\frac{1}{R^2}\int_{\hph}dV_{\hph}\,\psi(z)\overline{E_0(z;\tau,\xi)}\nonumber \\
&(\mc{K}^{-1}\tilde{\psi})(z)=\frac{2}{\pi^2}\int_0^{\infty}d\tau\,\tau\sinh(\pi\tau)\int_{\euc{}}d\xi\, \tilde{\psi}(\tau,\xi)E_0(z;\tau,\xi).
\end{align}
which is a combination of the Fourier Transform in $(x,\xi)$ variables and Kontorovich-Lebedev Transform in $(y,\tau)$ variables \cite{Leb}. The range of $\mc{K}$ can be formulated by considering $L^2(\euc{},d\xi)$ as the Hilbert Space corresponding to $\xi$, and then taking a direct integral of it with respect to the measure space $(0,\infty)$. So the map $\mc{K}$ can be expressed formally as
\begin{align}
\mc{K}:L^2(\hph,dV_{\hph})\rightarrow\int_{(0,\infty)}^{\oplus}L^2(\euc{},d\xi).
\end{align}
To prove that $\mc{K}$ provide a spectral representation of $\Delta_{\hph}$, we need two identities involving modified Bessel Functions which are given by \cite{His_1994}
\begin{align}\label{kl1}
\frac{2}{\pi^2}\int_{0}^{\infty}d\tau\,\tau\sinh(\pi\tau)K_{i\tau}(u)K_{i\tau}(v)=v\delta(u-v),
\end{align}
and
\begin{align}\label{kl2}
\frac{2}{\pi^2}\int_{0}^{\infty}\frac{du}{u}\,K_{i\tau}(u)K_{i\tau'}(u)=\frac{\delta(\tau-\tau')}{\sqrt{\tau\tau'}\sqrt{\sinh(\pi\tau)\sinh(\pi\tau')}}.
\end{align}
Moreover we shall also use the property $K_{i\tau}(u)=K_{-i\tau}(u)$. Using \eqref{kl2} one can show that $\mc{K}$ diagonalizes the Laplacian, that is for all $\tilde{\psi}\in \int_{(0,\infty)}^{\oplus}L^2(\euc{},d\xi)$ we have
\begin{align}\label{eigh}
-(\mc{K}\tilde{\Delta}_{\hph}\mc{K}^{-1}\tilde{\psi})(\tau,\xi)=\left(\tau^2+\frac{1}{4}\right)\tilde{\psi}(\tau,\xi),
\end{align}
and therefore
\begin{align}\label{eigh*}
-(\mc{K}\Delta_{\hph}\mc{K}^{-1}\tilde{\psi})(\tau,\xi)=\frac{1}{R^2}\left(\tau^2+\frac{1}{4}\right)\tilde{\psi}(\tau,\xi),
\end{align}
So the spectrum of $\Delta_{\hph}$ is given by
\begin{align}
\sigma(\Delta_{\hph})=\Big[\frac{1}{4R^2},\infty\Big).
\end{align}
By using \eqref{kl1} it is straightforward to show that $\mc{K}$ is an isomorphism, i.e. for all $\psi\in L^2(\hph,dV_{\hph})$ we have
\begin{align}
(\mc{K}^{-1}\mc{K}\psi)(z)=\psi(z).
\end{align}
Therefore, the transformation $\mc{K}$ with the eigenfunctions given as in \eqref{heig} provide a complete spectral representation of $\Delta_{\hph}$.
\subsection{Formulation of the Problem}
Just like the flat case we consider a particle of mass $m$ interacting with a Dirac-Delta potential on the hyperbolic plane $\hph$. Let $z_0=(x_0,y_0), y_0\neq 0$ denote the location of the Dirac-Delta potential. The corresponding Schrödinger equation for the bound state is written in coordinates $\hbar=1, 2m=1$ as
\begin{align}\label{eigh1}
\left(-\Delta_{\hph}-g\delta_{\hph}(z,z_0)\right)\phi(z)=-\nu^2\phi(z).
\end{align}
On a Riemannian manifold $(\mc{M},g)$ the Dirac Delta function $\delta_g(z,z_0)$ is defined such that
\begin{align}
\int_{\mc{M}}dV_g(z)\,\delta_g(z,z_0)=1\mtxt{for all}z_0\in \mc{M}.
\end{align}
Thus $\delta_{\hph}(z,z_0)$ is given by
\begin{align}
\delta_{\hph}(z,z_0)=\frac{y^2}{R^2}\delta(x,x_0)\delta(y,y_0).
\end{align}
Since $\mc{K}$ is an isomorphism we can write \eqref{eigh1} as
\begin{align}
-\mc{K}\Delta_{\hph}\mc{K}^{-1}\tilde{\phi}(\tau,\xi)-g\mc{K}\delta_{\hph}(z,z_0)\mc{K}^{-1}\tilde{\phi}(\tau,\xi)=-\nu^2\tilde{\phi}(\tau,\xi).
\end{align}
First term can be read directly from \eqref{eigh}, which is
\begin{align}\label{eigh2}
-\mc{K}\Delta_{\hph}\mc{K}^{-1}\tilde{\phi}(\tau,\xi)=\frac{1}{R^2}\left(\tau^2+\frac{1}{4}\right)\tilde{\phi}(\tau,\xi).
\end{align}
Second term can be easily calculated using $\delta_{\hph}(z,z_0)$. The result is
\begin{align}\label{eigh3}
g\mc{K}\delta_{\hph}(z,z_0)\mc{K}^{-1}\tilde{\phi}(\tau,\xi)=\frac{2g}{\pi^2R^2}&\int_0^{\infty}d\tau'\,\tau'\sinh(\pi\tau')\int_{\euc{}}d\xi'\,\overline{E_0(z_0,\tau,\xi)}E_0(z_0,\tau',\xi')\tilde{\phi}(\tau',\xi')
\end{align}
If we put \eqref{eigh2} and \eqref{eigh3} into the \eqref{eigh1} and rearrange the terms we obtain
\begin{align}\label{eigh4}
\left(\tau^2+a^2\right)\tilde{\phi}(\tau,\xi)=\frac{2g}{\pi^2}&\int_0^{\infty}d\tau'\,\tau'\sinh(\pi\tau')\int_{\euc{}}d\xi'\,\overline{E_0(z_0,\tau,\xi)}E_0(z_0,\tau',\xi')\tilde{\phi}(\tau',\xi'),
\end{align}
where we made the definition $a^2 \equiv \frac{1}{4}+\nu^2R^2$. Our next goal is to determine the type and the cause of the divergence. To this end we make an attempt to solve \eqref{eigh4}. We define
\begin{align}\label{eigh5}
\mc{N}\equiv \int_0^{\infty}d\tau'\,\tau'\sinh(\pi\tau')\int_{\euc{}}d\xi'\,E_0(z_0,\tau',\xi')\tilde{\phi}(\tau',\xi').
\end{align}
Then $\tilde{\phi}(\tau,\xi)$ becomes
\begin{align}
\tilde{\phi}(\tau,\xi)=\frac{2g}{\pi^2}\mc{N}\overline{E_0(z_0,\tau,\xi)}\left(\tau^2+a^2\right)^{-1}.
\end{align}
We put this result back into \eqref{eigh5} and find
\begin{align}\label{eigh6}
\frac{1}{g}=\frac{2}{\pi^2}\int_0^{\infty}d\tau'\,\tau'\sinh(\pi\tau')\left(\tau'^2+a^2\right)^{-1}\int_{\euc{}}d\xi'\,\overline{E_0(z_0,\tau',\xi')}E_0(z_0,\tau',\xi').
\end{align}
Let us denote the $\xi$-integral by $\Upsilon(\tau')$. Using the explicit form of the eigenfunctions as given in \eqref{heig} we can write $\Upsilon(\tau')$ as
\begin{align}
\Upsilon(\tau')&=\frac{y_0}{2\pi}\int_{\euc{}}d\xi'\,K_{i\tau'}(y_0\abs{\xi})K_{-i\tau'}(y_0\abs{\xi})\nonumber \\
&=\frac{y_0}{\pi}\int_{0}^{\infty}d\xi'\,K_{i\tau'}(y_0\xi)K_{i\tau'}(y_0\xi)
\end{align}
To evaluate the integral we will use the following integral representation of the modified Bessel functions \cite{Arfken}:
\begin{align}
K_{\nu}(z)=\int_0^{\infty}du\,e^{-z\cosh u}\cosh(\nu u),\quad \textrm{Re}\,z>0
\end{align}
Therefore $\Upsilon(\tau')$ becomes
\begin{align}
\Upsilon(\tau') &=\frac{y_0}{\pi}\int_0^{\infty}d\xi\int_0^{\infty}du\int_0^{\infty}dv\,e^{-y_0\xi(\cosh u+\cosh v)}\cosh(i\tau' u)\cosh(i\tau' v)\nonumber \\
&=\frac{y_0}{\pi}\int_0^{\infty}du\int_0^{\infty}dv \cos(\tau' u)\cos(\tau' v)\int_0^{\infty}d\xi \,e^{-y_0\xi(\cosh u+\cosh v)}\nonumber \\
&=\frac{1}{\pi}\int_0^{\infty}du\,\cos(\tau' u)\int_0^{\infty}dv \frac{\cos(\tau' v)}{\cosh u+\cosh v}.
\end{align}
The $dv$ integral can be evaluated using the definite integral \cite{Integral}
\begin{align}\label{def1}
\int_0^{\infty}\frac{\cos(ax)dx}{b\cosh(\beta x)+c}=\frac{\pi\sin\left(\frac{a}{\beta}\cosh^{-1}\left(\frac{c}{b}\right)\right)}{\beta\sqrt{c^2-b^2}\sinh\left(\frac{a\pi}{\beta}\right)},\quad \mtxt{for} c>b>0.
\end{align}
In our case $a=\tau'$, $b=1$, $\beta=1$, $c=\cosh u$ and $\cosh u\geq 1>0,$ for all $u\in[0,\infty)$ so we can use \eqref{def1}. Hence $dv$ integral becomes
\begin{align}
\int_0^{\infty}dv \frac{\cos(\tau' v)}{\cosh u+\cosh v}=\frac{\pi\sin\left(\tau' \cosh^{-1}(\cosh u)\right)}{\sqrt{\cosh^2 -1}\sinh(\tau' \pi)}=\frac{\pi \sin(\tau' u)}{\sinh(u)\sinh(\tau' \pi)}.
\end{align}
Then we have
\begin{align}
\Upsilon(\tau') = \frac{1}{\sinh(\tau' \pi)}\int_0^{\infty}du\,\frac{\cos(\tau' u)\sin(\tau' u)}{\sinh u}.
\end{align}
Finally we will use \cite{Integral}
\begin{align}
\int_0^{\infty}dx\,\frac{\sin(\alpha x)\cos(\beta x)}{\sinh (\gamma x)}=\frac{\pi\sinh\left(\frac{\pi a}{\gamma}\right)}{2\gamma \left(\cosh\left(\frac{\alpha \pi}{\gamma}\right)+\cosh\left(\frac{\beta \pi}{\gamma}\right)\right)}
\end{align}
for $\textrm{Im}\,(\alpha+\beta)<\textrm{Re}\,\gamma$. In our case $\alpha=\beta=\tau'$, $\gamma=1$ and $\tau'\in \mathbb{R}$, therefore $ \text{Im}\,(\alpha+\beta)=0<1$. Thus
\begin{align}
\int_0^{\infty}du\,\frac{\cos(\tau' u)\sin(\tau' u)}{\sinh u}=\frac{\pi\sinh(\pi\tau')}{2\left(\cosh(\pi\tau')+\cosh(\pi\tau')\right)}=\frac{\pi}{4}\tanh(\pi\tau')
\end{align}
and $\Upsilon(\tau')$ becomes
\begin{align}\label{ups}
\Upsilon(\tau')=\int_{\euc{}}d\xi'\,\overline{E_0(z_0,\tau',\xi')}E_0(z_0,\tau',\xi')=\frac{\pi}{4}\frac{\tanh(\pi\tau')}{\sinh(\pi\tau')}.
\end{align}
Finally we put this result into \eqref{eigh6} and obtain
\begin{align}\label{hph_fin}
\frac{1}{g}=\frac{1}{2\pi}\int_0^{\infty}d\tau'\,\tau'\tanh(\pi\tau')\left(\tau'^2+a^2\right)^{-1}.
\end{align}
For large values of $\tau'$, $\tanh(\pi\tau')\approx 1$ and the integrand behaves as $1/\tau'$ so just like the flat case we face with a logarithmic divergence. This analysis also shows us that there is no divergence in the $\xi$ term. Therefore we only need to concern with the renormalization of $\tau$.
\subsection{Applying the ERG Procedure}
We start by writing the eigenvalue equation at the bare scale $\Lambda$.
\begin{align}\label{herg1}
\left(\tau^2+a^2\right)\tilde{\phi}(\tau,\xi)=\frac{2}{\pi^2}\Theta_{\Lambda}(\tau)\int_0^{\Lambda}d\tau'\,g_{\Lambda}(\tau,\tau')\tau'\sinh(\pi\tau')\vartheta(\tau,\tau';\xi),
\end{align}
where $g_{\Lambda}(\tau,\tau')=g-x_{\Lambda}(\tau,\tau')$ and
\begin{align}\label{herg2}
\vartheta(\tau,\tau';\xi)\equiv \int_{\euc{}}d\xi'\,\overline{E_0(z_0,\tau,\xi)}E_0(z_0,\tau',\xi')\tilde{\phi}(\tau',\xi').
\end{align}
At an infinitesimally lower scale $\Lambda-d\Lambda$ we write
\begin{align}\label{herg3}
\left(\tau^2+a^2\right)\tilde{\phi}(\tau,\xi)=\frac{2}{\pi^2}\Theta_{\Lambda-d\Lambda}(\tau)\int_0^{\Lambda-d\Lambda}d\tau'\,g_{\Lambda-d\Lambda}(\tau,\tau')\tau'\sinh(\pi\tau')\vartheta(\tau,\tau';\xi).
\end{align}
We can rewrite \eqref{herg1} as
\begin{align}\label{herg4}
\left(\tau^2+a^2\right)\tilde{\phi}(\tau,\xi)=\frac{2}{\pi^2}\Theta_{\Lambda}(\tau)\Bigg(&\int_0^{\Lambda-d\Lambda}d\tau'\,g_{\Lambda}(\tau,\tau')\tau'\sinh(\pi\tau')\vartheta(\tau,\tau';\xi)\nonumber \\
&\quad+d\Lambda\,g_{\Lambda}(\tau,\Lambda)\Lambda\sinh(\pi\Lambda)\vartheta(\tau,\Lambda;\xi)\Bigg),
\end{align}
and for $\tau=\Lambda$ we obtain
\begin{align}\label{herg5}
\left(\Lambda^2+a^2\right)\tilde{\phi}(\Lambda,\xi)=\frac{2}{\pi^2}\Bigg(&\int_0^{\Lambda-d\Lambda}d\tau'\,g_{\Lambda}(\Lambda,\tau')\tau'\sinh(\pi\tau')\vartheta(\Lambda,\tau';\xi)\nonumber \\
&\quad+d\Lambda\,g_{\Lambda}(\Lambda,\Lambda)\Lambda\sinh(\pi\Lambda)\vartheta(\Lambda,\Lambda;\xi)\Bigg).
\end{align}
Then $\tilde{\phi}(\Lambda,\xi)$ becomes
\begin{align}\label{herg6}
\tilde{\phi}(\Lambda,\xi)=\frac{2}{\pi^2}\left(\Lambda^2+a^2\right)^{-1}\int_0^{\Lambda-d\Lambda}d\tau'\,g_{\Lambda}(\Lambda,\tau')\tau'\sinh(\pi\tau')\vartheta(\Lambda,\tau';\xi),
\end{align}
where we have again ignored the term proportional to $d\Lambda$. From this result we can find $\vartheta(\tau,\Lambda;\xi)$ as
\begin{align}
\vartheta(\tau,\Lambda;\xi)=&\frac{2}{\pi^2}\left(\Lambda^2+a^2\right)^{-1}\int_0^{\Lambda-d\Lambda}d\tau'\,g_{\Lambda}(\Lambda,\tau')\tau'\sinh(\pi\tau')\nonumber \\ &\times \int_{\euc{}}d\xi'\,\overline{E_0(z_0,\tau,\xi)}E_0(z_0,\Lambda,\xi')\vartheta(\Lambda,\tau';\xi')
\end{align}
By putting the explicit expression for $\vartheta(\Lambda,\tau';\xi')$ we get
\begin{align}
\vartheta(\tau,\Lambda;\xi)=&\frac{2}{\pi^2}\left(\Lambda^2+a^2\right)^{-1}\int_0^{\Lambda-d\Lambda}d\tau'\,g_{\Lambda}(\Lambda,\tau')\tau'\sinh(\pi\tau')\nonumber \\ 
&\times \int_{\euc{}}d\xi'\,\overline{E_0(z_0,\tau,\xi)}E_0(z_0,\Lambda,\xi')\int_{\euc{}}d\xi''\,\overline{E_0(z_0,\Lambda,\xi')}E_0(z_0,\tau',\xi'')\tilde{\phi}(\tau',\xi'')\nonumber \\
&=\frac{2}{\pi^2}\left(\Lambda^2+a^2\right)^{-1}\int_0^{\Lambda-d\Lambda}d\tau'\,g_{\Lambda}(\Lambda,\tau')\tau'\sinh(\pi\tau')\nonumber \\ 
&\times \int_{\euc{}}d\xi'\,\overline{E_0(z_0,\Lambda,\xi')}E_0(z_0,\Lambda,\xi') \int_{\euc{}}d\xi''\,\overline{E_0(z_0,\tau,\xi)}E_0(z_0,\tau',\xi'')\tilde{\phi}(\tau',\xi'')\nonumber \\
&=\frac{1}{2\pi}\left(\Lambda^2+a^2\right)^{-1}\frac{\tanh(\pi\Lambda)}{\sinh(\pi\Lambda)}\int_0^{\Lambda-d\Lambda}d\tau'\,g_{\Lambda}(\Lambda,\tau')\tau'\sinh(\pi\tau')\vartheta(\tau,\tau';\xi).
\end{align}
If we put this result back into \eqref{herg4} we find
\begin{align}\label{herg7}
\left(\tau^2+a^2\right)\tilde{\phi}(\tau,\xi)=\frac{2\Theta_{\Lambda}(\tau)}{\pi^2}&\int_0^{\Lambda-d\Lambda}\!\!\!\!d\tau'\tau'\sinh(\pi\tau')\vartheta(\tau,\tau';\xi)\nonumber \\
&\times\Bigg(g_{\Lambda}(\tau,\tau')+\frac{\Lambda\tanh(\pi\Lambda)}{2\pi(\Lambda^2+a^2)}g_{\Lambda}(\Lambda,\tau')g_{\Lambda}(\tau,\Lambda)\Bigg)
\end{align}
By comparing this with \eqref{herg3} we arrive to an equation for the coupling constant
\begin{align}
g_{\Lambda-d\Lambda}(\tau,\tau')=g_{\Lambda}(\tau,\tau')+\frac{\Lambda\tanh(\pi\Lambda)}{2\pi(\Lambda^2+a^2)}g_{\Lambda}(\Lambda,\tau')g_{\Lambda}(\tau,\Lambda),
\end{align}
which can be put into differential form as
\begin{align}
-\d{g_{\Lambda}(\tau,\tau')}{\Lambda}=\frac{\Lambda\tanh(\pi\Lambda)}{2\pi(\Lambda^2+a^2)}g_{\Lambda}(\Lambda,\tau')g_{\Lambda}(\tau,\Lambda),
\end{align}
and by integrating from $\lambda$ to $\Lambda$ we find
\begin{align}\label{herg7_1}
g_{\lambda}(\tau,\tau')=g-x_{\Lambda}(\tau,\tau')+\frac{1}{2\pi}\int_{\lambda}^{\Lambda}ds\,\frac{s}{s^2+a^2}\tanh(\pi s)g_{s}(s,\tau')g_{s}(\tau,s).
\end{align}
To obtain a solution we shall use the same procedure we used in the flat case. We begin with
\begin{align}
g_{\lambda}^{(1)}(\tau,\tau')=g\mtxt{so that}x_{\Lambda}^{(1)}(\tau,\tau')=0
\end{align}
Then $g_{\lambda}^{(2)}$ becomes
\begin{align}
g_{\lambda}^{(2)}=g-x_{\Lambda}^{(2)}(\tau,\tau')+\frac{g^2}{2\pi}\int_{\lambda}^{\Lambda}ds\,\frac{s\tanh(\pi s)}{s^2+a^2}.
\end{align}
We choose the counterterm as
\begin{align}
x_{\Lambda}^{(2)}(\tau,\tau')=\frac{g^2}{2\pi}\int_{\lambda_0}^{\Lambda}ds\,\frac{s\tanh(\pi s)}{s^2+a^2},
\end{align}
so that the effective coupling at the second order is now finite and given by
\begin{align}
g_{\lambda}^{(2)}(\tau,\tau')=g-\frac{g^2}{2\pi}\int_{\lambda_0}^{\lambda}ds\,\frac{s\tanh(\pi s)}{s^2+a^2}.
\end{align}
Just like the flat case, $g_{\lambda}^{(n)}(\tau,\tau')$ is independent of $\tau$ and $\tau'$ for all $n$. At the order $n+1$, the effective coupling becomes
\begin{align}
g_{\lambda}^{(n+1)}=g-x_{\Lambda}^{(n+1)}+\frac{1}{2\pi}\int_{\lambda}^{\Lambda}ds\,\frac{s\tanh(\pi s)}{s^2+a^2}\left(g_s^{(n)}\right)^2.
\end{align}
By choosing the counterterm as
\begin{align}
x_{\Lambda}^{(n+1)}=\frac{1}{2\pi}\int_{\lambda_0}^{\Lambda}ds\,\frac{s\tanh(\pi s)}{s^2+a^2}\left(g_s^{(n)}\right)^2,
\end{align}
we find
\begin{align}
g_{\lambda}^{(n+1)}=g-\frac{1}{2\pi}\int_{\lambda_0}^{\lambda}ds\,\frac{s\tanh(\pi s)}{s^2+a^2}\left(g_s^{(n)}\right)^2.
\end{align}
If we assume the existence of
\begin{align*}
\lim_{n\rightarrow\infty}g_{\lambda}^{(n+1)}=g_{\lambda},
\end{align*}
we can write for the effective coupling
\begin{align}
g_{\lambda}=g-\frac{1}{2\pi}\int_{\lambda_0}^{\lambda}ds\,\frac{s\tanh(\pi s)}{s^2+a^2}g_s^2,
\end{align}
which again implies $g=g_{\lambda_0}$. This expression can be put into the following form
\begin{align}
-\int_{\lambda_0}^{\lambda}dg_s\,\frac{dg_s}{g_s^2}=\frac{1}{2\pi}\int_{\lambda_0}^{\lambda}ds\,\frac{s\tanh(\pi s)}{s^2+a^2}.
\end{align}
By evaluating the integral on the LHS and expressing $\tanh(\pi s)$ as
\begin{align*}
\tanh(\pi s)=1-\frac{2}{e^{2\pi s}+1},
\end{align*}
we obtain
\begin{align}\label{hfs2}
\frac{1}{g_{\lambda}}=\frac{1}{g_{\lambda_0}}+\frac{1}{2\pi}\int_{\lambda_0}^{\lambda}ds\,\frac{s}{s^2+a^2}-\frac{1}{2\pi}\int_{\lambda_0}^{\lambda}ds\,\frac{s}{s^2+a^2}\frac{2}{e^{2\pi s}+1}.
\end{align}
Let us define a function $\alpha(s)$ by
\begin{align}
\alpha^{-1}(s)\equiv \frac{1}{2\pi}\int_{0}^{s}ds'\,\frac{s'}{{s'}^2+a^2}\frac{2}{e^{2\pi s'}+1}
\end{align}
Then \eqref{hfs2} becomes
\begin{align}\label{hfs3}
\frac{1}{g_{\lambda}}=\frac{1}{g_{\lambda_0}}+\frac{1}{2\pi}\int_{\lambda_0}^{\lambda}ds\,\frac{s}{s^2+a^2}-\frac{1}{\alpha(\lambda)}+\frac{1}{\alpha(\lambda_0)}.
\end{align}
By redefining the coupling constant by $\tilde{g}^{-1}_{s}\equiv g^{-1}_s+\alpha^{-1}_{s}$ and evaluating the integral in \eqref{hfs3} we obtain
\begin{align}
\frac{1}{\tilde{g}_{\lambda}}=\frac{1}{\tilde{g}_{\lambda_0}}+\frac{1}{4\pi}\log\left(\frac{\lambda^2+a^2}{\lambda_0^2+a^2}\right),
\end{align}
which can be solved as
\begin{align}
\tilde{g}_{\lambda}=\frac{\tilde{g}_{\lambda_0}}{1+\frac{\tilde{g}_{\lambda_0}}{4\pi}\log\left(\frac{\lambda^2+a^2}{\lambda_0^2+a^2}\right)}.
\end{align}
We see that by slightly modifying the coupling constant we can obtain the same solution as the flat case. To show that this modification brings no problems at high energies let us try to estimate $\alpha^{-1}(\lambda)$.
\begin{align}
\alpha^{-1}(\lambda)=\frac{1}{2\pi}\int_{0}^{\lambda}ds\,\frac{s}{s^2+a^2}\frac{2}{e^{2\pi s}+1}\leq \frac{1}{\pi}\int_{0}^{\lambda}ds\,\frac{se^{-2\pi s}}{s^2+a^2}
\end{align}
Using Cauchy-Schwartz inequality we find
\begin{align}
\alpha^{-1}(\lambda) &\leq \frac{1}{\pi}\left[\int_{0}^{\lambda}ds\,\frac{s^2}{(s^2+a^2)^2}\right]^{1/2}\left[\int_{0}^{\lambda}ds\,e^{-4\pi s}\right]^{1/2}\nonumber \\
&=\frac{1}{\pi}\left[\frac{1}{2}\left(\frac{\tan^{-1}\left(\frac{\lambda}{a}\right)}{a}-\frac{\lambda}{\lambda^2+a^2}\right)\right]^{1/2}\left[\frac{1}{4\pi}\left(1-e^{-4\pi\lambda}\right)\right]^{1/2}.
\end{align}
We see that for high energies $\lambda \gg 1$, this correction term behaves like a constant.

\subsection{Estimating the Range of Renormalizability}
Investigating that under which conditions does the sequence $\lbrace g_{\lambda}^{(n)}\rbrace$ has a limit is identical to what we have done for the flat case in Section \ref{cont_r}. In this case we define the map $T:C(I)\rightarrow C(I)$ as
\begin{align}
T(\mathfrak{g})(\lambda)=g_{\lambda_0}-\frac{1}{2\pi}\int_{\lambda_0}^{\lambda}ds\,\frac{s\,\tanh(\pi s)}{s^2+a^2}g_s^2
\end{align}
Then we have the estimate
\begin{align}
\abs{T(\mathfrak{g})-T(\mathfrak{h})}&=\frac{1}{2\pi}\int_{\lambda_0}^{\lambda}ds\,\frac{s\,\tanh(\pi s)}{s^2+\nu^2}\left((h_s)^2-(g_s)^2\right)\nonumber \\
&\leq\frac{1}{2\pi}\sup_{s\in [\lambda_0,\lambda]}\abs{(h_s)^2-(g_s)^2}\int_{\lambda_0}^{\lambda}ds\,\frac{s\,\tanh(\pi s)}{s^2+\nu^2}\nonumber \\
&\leq\frac{1}{2\pi}\sup_{s\in [\lambda_0,\lambda]}\abs{(h_s)^2-(g_s)^2}\int_{\lambda_0}^{\lambda}ds\,\frac{1}{s}\nonumber \\
&=\frac{1}{2\pi}\sup_{s\in [\lambda_0,\lambda]}\abs{(h_s+g_s)(h_s-g_s)}\log\left(\frac{\lambda}{\lambda_0}\right)
\end{align}
Just like the flat case, $T$ is a contraction if
\begin{align}
\frac{g_{\lambda_0}}{\pi}\log\left(\frac{\tilde{\lambda}}{\lambda_0}\right)<1.
\end{align}
\subsection{Bound State Solution}
To find the bound state energy we rewrite \eqref{hph_fin} in terms of renormalized coupling.
\begin{align}\label{bsh1}
\frac{1}{g_{\Lambda}}=\frac{1}{2\pi}\int_0^{\Lambda}d\tau'\,\tau'\tanh(\pi\tau')\left(\tau'^2+a^2\right)^{-1},
\end{align}
where
\begin{align}
\frac{1}{g_{\Lambda}}=\frac{1}{\tilde{g}_{\Lambda}}-\frac{1}{\alpha(\Lambda)}=\frac{1}{\tilde{g}_{\lambda_0}}+\frac{1}{4\pi}\log\left(\frac{\Lambda^2+a^2}{\lambda_0^2+a^2}\right)-\frac{1}{2\pi}\int_{0}^{\Lambda}ds\,\frac{s}{s^2+a^2}\frac{2}{e^{2\pi s}+1}.
\end{align}
Thus \eqref{bsh1} becomes
\begin{align}
\frac{1}{\tilde{g}_{\lambda_0}}+\frac{1}{4\pi}\log\left(\frac{\Lambda^2+a^2}{\lambda_0^2+a^2}\right)-\frac{1}{2\pi}\int_{0}^{\Lambda}ds\,\frac{s}{s^2+a^2}\frac{2}{e^{2\pi s}+1}=\frac{1}{2\pi}\int_0^{\Lambda}d\tau'\,\frac{\tau'}{{\tau'}^2+a^2}-\frac{1}{2\pi}\int_{0}^{\Lambda}d\tau'\,\frac{\tau'}{{\tau'}^2+a^2}\frac{2}{e^{2\pi \tau'}+1}
\end{align}
The integrals on both sides vanish so we are left with
\begin{align}
\frac{1}{\tilde{g}_{\lambda_0}}+\frac{1}{4\pi}\log\left(\frac{\Lambda^2+a^2}{\lambda_0^2+a^2}\right)=\frac{1}{4\pi}\log\left(\frac{\Lambda^2+a^2}{a^2}\right),
\end{align}
with the solution in the $\Lambda\rightarrow \infty$ limit
\begin{align}
\lim_{\Lambda\rightarrow \infty}-\nu^2R^2=-\lambda_0^2\frac{e^{-4\pi/\tilde{g}_{\lambda_0}}}{1-e^{-4\pi/\tilde{g}_{\lambda_0}}}+\frac{1}{4}.
\end{align}
\section{Exact Renormalization Group on the $2$-sphere}
\subsection{Formulation of the Problem}
Considering the same problems as in the previous sections we write the eigenvalue equation for the bound state as
\begin{align}\label{fors1}
(-\Delta_{\sph{2}}+\nu^2)\phi(\Omega)=g\delta_{\sph{2}}(\Omega,\Omega_0)\phi(\Omega),
\end{align}
where $\Omega_0\in \sph{2}$ is the location of the Dirac-Delta potential. Since the spherical harmonics $Y_{l}^{m}(\Omega)$ form a complete set, we can expand $\phi(\Omega)$ in terms of them and using the eigenvalue relation 
\begin{align*}
-\Delta_{\sph{2}}Y_l^m(\Omega)=R^{-2}l(l+1)Y_l^m(\Omega),
\end{align*}
we find
\begin{align}
\sum_{l=0}^{\infty}\sum_{m=-l}^{l}C_l^m\left[R^{-2}l(l+1)+\nu^2\right]Y_l^m(\Omega)=g\delta_{\sph{2}}(\Omega,\Omega_0)\sum_{l=0}^{\infty}\sum_{m=-l}^{l}C_l^mY_l^m(\Omega).
\end{align}
Now if we multiply both sides by $\overline{Y_{l'}^{m'}(\Omega)}$ and integrate over $R^{-2}\int_{\sph{2}}dV_{\sph{2}}$ we obtain
\begin{align}\label{fors2}
C_l^m\left[l(l+1)+\nu^2R^2\right]=g\sum_{l'=0}^{\infty}\sum_{m'=-l'}^{l'}C_{l'}^{m'}\overline{Y_l^m(\Omega_0)}Y_{l'}^{m'}(\Omega_0),
\end{align}
where we have also used the orthogonality relation of the spherical harmonics. The next step is to determine the type of divergence. For this we define
\begin{align}\label{fors3}
\mc{N}=\sum_{l'=0}^{\infty}\sum_{m'=-l'}^{l'}C_{l'}^{m'}Y_{l'}^{m'}(\Omega_0),
\end{align}
so that $C_l^m$ is given by
\begin{align}
C_l^m=\mc{N}\frac{g\overline{Y_l^m(\Omega_0)}}{l(l+1)+\nu^2R^2}.
\end{align}
By plugging this result into \eqref{fors3} we find $g^{-1}$ as
\begin{align}\label{fors4}
\frac{1}{g}=\frac{1}{4\pi}\sum_{l'=0}^{\infty}\frac{2l'+1}{l'(l'+1)+\nu^2R^2},
\end{align}
where we have used
\begin{align}
\sum_{m'=-l'}^{l'}\overline{Y_{l'}^{m'}(\Omega_0)}Y_{l'}^{m'}(\Omega_0)=\frac{2l'+1}{4\pi}.
\end{align}
By using Maclaurin-Cauchy integral test we can see that the $l'$ sum in \eqref{fors4} is logarithmically divergent and the divergence is caused by the large $l'$ values. 
\subsection{Applying the ERG Procedure}
We begin by writing the eigenvalue equation at the bare scale $\Lambda$.
\begin{align}\label{ergs1}
C_l^m\left[l(l+1)+\nu^2R^2\right]=\Theta_{\Lambda}(l)\sum_{l'=0}^{\Lambda}g_{\Lambda}(l,l')\vartheta(l,l';m),
\end{align}
where
\begin{align}\label{ergs2}
\vartheta(l,l';m)\equiv\sum_{m'=-l'}^{l'}C_{l'}^{m'}\overline{Y_l^m(\Omega_0)}Y_{l'}^{m'}(\Omega_0).
\end{align}
Since the eigenvalue spectrum is discrete, we take the second cutoff as $\Lambda-1$ instead of $\Lambda-d\Lambda$. We write
\begin{align}\label{ergs3}
C_l^m\left[l(l+1)+\nu^2R^2\right]=\Theta_{\Lambda-1}(l)\sum_{l'=0}^{\Lambda-1}g_{\Lambda-1}(l,l')\vartheta(l,l';m).
\end{align}
We rewrite \eqref{ergs1} as
\begin{align}\label{ergs4}
C_l^m\left[l(l+1)+\nu^2R^2\right]=\Theta_{\Lambda}(l)\left(\sum_{l'=0}^{\Lambda-1}g_{\Lambda}(l,l')\vartheta(l,l';m)+g_{\Lambda}(l,\Lambda)\vartheta(l,\Lambda;m)\right)
\end{align}
so that by substituting $l=\Lambda$ we get an expression for $C_{\Lambda}^{m}$:
\begin{align}\label{ergs5}
C_{\Lambda}^{m}=\frac{1}{\Lambda(\Lambda+1)+\nu^2R^2}\left(\sum_{l'=0}^{\Lambda-1}g_{\Lambda}(\Lambda,l')\vartheta(\Lambda,l';m)+g_{\Lambda}(\Lambda,\Lambda)\vartheta(\Lambda,\Lambda;m)\right).
\end{align}
We note that due to the discrete spectrum we could not ignore the second term. Using \eqref{ergs2} and \eqref{ergs5} we can write
\begin{align}
\vartheta(l,\Lambda;m)&=\sum_{m'=-\Lambda}^{\Lambda}C_{\Lambda}^{m'}Y_{\Lambda}^{m'}(\Omega_0)\overline{Y_l^m(\Omega_0)}\nonumber\\
&=\sum_{m'=-\Lambda}^{\Lambda}\frac{Y_{\Lambda}^{m'}(\Omega_0)\overline{Y_l^m(\Omega_0)}}{\Lambda(\Lambda+1)+\nu^2R^2}\left(\sum_{l'=0}^{\Lambda-1}g_{\Lambda}(\Lambda,l')\vartheta(\Lambda,l';m)+ g_{\Lambda}(\Lambda,\Lambda)\vartheta(\Lambda,\Lambda;m)\right)\nonumber\\
&\hspace{1.5in}+g_{\Lambda}(\Lambda,\Lambda)\sum_{m'=-\Lambda}^{\Lambda}Y_{\Lambda}^{m'}(\Omega_0)\overline{Y_l^m(\Omega_0)}\vartheta(\Lambda,\Lambda;m')\Bigg).
\end{align}
By putting explicit expressions for $\vartheta(\Lambda,l';m')$ and $\vartheta(\Lambda,\Lambda;m')$ we get
\begin{align}
\vartheta(l,\Lambda;m)&=\frac{1}{\Lambda(\Lambda+1)+\nu^2R^2}\Bigg(\sum_{l'=0}^{\Lambda-1}g_{\Lambda}(\Lambda,l')\sum_{m'=-\Lambda}^{\Lambda}Y_{\Lambda}^{m'}(\Omega_0)\overline{Y_{\Lambda}^{m'}(\Omega_0)}\nonumber \\ &\hspace{1.3in}\times\sum_{m''=-l'}^{l'}C_{l'}^{m''}Y_{l'}^{m''}(\Omega_0)\overline{Y_l^m(\Omega_0)}\nonumber\\
&\hspace{1.3in}+g_{\Lambda}(\Lambda,\Lambda)\sum_{m'=-\Lambda}^{\Lambda}Y_{\Lambda}^{m'}(\Omega_0)\overline{Y_{\Lambda}^{m'}(\Omega_0)}\nonumber \\
&\hspace{1.3in}\times\sum_{m''=-\Lambda}^{\Lambda}C_{\Lambda}^{m''}Y_{\Lambda}^{m''}(\Omega_0)\overline{Y_l^m(\Omega_0)}\Bigg)\nonumber \\
&=\frac{1}{4\pi}\frac{2\Lambda +1}{\Lambda(\Lambda +1)+\nu^2R^2}\left(\sum_{l'=0}^{\Lambda -1}g_{\Lambda}(\Lambda, l')\vartheta(l,l';m)+g_{\Lambda}(\Lambda,\Lambda)\vartheta(l,\Lambda;m)\right)
\end{align}
From this, $\vartheta(l,\Lambda;m)$ can be solved as
\begin{align}
\vartheta(l,\Lambda;m)=\left(4\pi\frac{\Lambda(\Lambda+1)+\nu^2R^2}{2\Lambda+1}-g_{\Lambda}(\Lambda,\Lambda)\right)^{-1}\;\sum_{l'=0}^{\Lambda -1}g_{\Lambda}(\Lambda, l')\vartheta(l,l';m).
\end{align}
By putting this result back into \eqref{ergs4} we get
\begin{align}
C_l^m[l(l+1)+\nu^2R^2]&=\Theta_{\Lambda}(l)\sum_{l'=0}^{\Lambda-1}\Bigg[g_{\Lambda}(l,l')+\left(4\pi\frac{\Lambda(\Lambda+1)-M}{2\Lambda+1}-g_{\Lambda}(\Lambda,\Lambda)\right)^{-1}\nonumber \\
&\hspace{1in}\times g_{\Lambda}(l,\Lambda)g_{\Lambda}(\Lambda, l')\Bigg]\vartheta(l,l';m),
\end{align}
and by comparing this result with \eqref{ergs3} we obtain a recursion relation for the effective coupling constant.
\begin{align}
g_{\Lambda-1}(l,l')=g_{\Lambda}(l,l')+\left(4\pi\frac{\Lambda(\Lambda+1)+\nu^2R^2}{2\Lambda+1}-g_{\Lambda}(\Lambda,\Lambda)\right)^{-1}g_{\Lambda}(l,\Lambda)g_{\Lambda}(\Lambda, l')
\end{align}
From this relation we can express the effective coupling at the effective scale $\lambda$ as
\begin{align}
g_{\lambda}(l,l')=g_{\Lambda}(l,l')+\sum_{s=\lambda+1}^{\Lambda}\left(4\pi\frac{s(s+1)+\nu^2R^2}{2s+1}-g_{s}(s,s)\right)^{-1}g_{s}(l,s)g_{s}(s, l'),
\end{align}
or
\begin{align}
g_{\lambda}(l,l')=g-x_{\Lambda}(l,l')+\frac{1}{4\pi}\sum_{s=\lambda+1}^{\Lambda}\left(\frac{s(s+1)+\nu^2R^2}{2s+1}-\frac{g_{s}(s,s)}{4\pi}\right)^{-1}g_{s}(l,s)g_{s}(s, l').
\end{align}
By applying the same iteration procedure as in the previous cases we can obtain
\begin{align}
g_{\lambda}^{(n+1)}=g-\frac{1}{4\pi}\sum_{s=\lambda_0}^{\lambda}\left(\frac{s(s+1)+\nu^2R^2}{2s+1}-\frac{g_{s}^{(n)}}{4\pi}\right)^{-1}(g_{s}^{(n)})^2.
\end{align}
This is a very complicated recursive relation, and unlike previous cases, we could not convert it to a differential equation, from which we can solve for $g_{\lambda}$ in the $n\rightarrow \infty$ limit.
\section{Conclusion}
In this paper we investigated a non-perturbative renormalization of point interactions on the two-dimensional hyperbolic space, using the Exact Renormalization Group method. We showed that the theory is asymptotically free and the flow equations has the same form as in the flat case. 
\section*{Acknowledgments}
O. T. Turgut would like to thank Prof. P. Exner and Prof. M. Znojil for the kind invitation to AAMP XI meeting which was held in Villa Lanna, and for the kind support for his lodging there.
\bibliographystyle{plain}
\bibliography{references}
\end{document}